\documentclass[a4paper]{jpconf}
\usepackage[dvipsnames]{xcolor}
\usepackage{graphicx}
\usepackage{url}
\usepackage{amsmath}
\usepackage{amssymb}
\usepackage{latexsym}
\usepackage{citesort}

\newcommand\Comment[1]{{\vspace{5mm}\noindent\bfseries\color{NavyBlue}[#1]\\}}
\renewcommand{\Comment}[1]{}

\begin{document}

\title{Track 3: Computations in theoretical physics -- techniques and methods}
\author{Gionata Luisoni$^1$, Stanislav Poslavsky$^2$ and York Schr\"oder$^3$}
\address{$^1$ Theoretical Physics Department, CERN, Geneva, Switzerland}
\address{$^2$ Institute for High Energy Physics NRC ``Kurchatov Institute'', 142281 Protvino, Russia}
\address{$^3$ Grupo de Fisica de Altas Energias, Universidad del Bio-Bio, Casilla 447, Chillan, Chile}
\ead{gionata.luisoni@cern.ch, stvlpos@mail.ru, yschroeder@ubiobio.cl}

\begin{abstract}
Here, we attempt to summarize the activities of Track 3 of the 17th
International Workshop on Advanced Computing and Analysis Techniques
in Physics Research (ACAT 2016).\\[2mm] \mbox{}\hfill {CERN-TH-2016-081}
\end{abstract}

\section{Introduction}

Today's computations in theoretical high energy physics (HEP) are
hardly feasible without the use of modern computer tools and advanced
computational algorithms. This is due to the immense progress in both
theoretical and experimental techniques that we have witnessed in
recent decades, especially with the launch of powerful machines such
as the LHC. As a matter of fact, the processes of interest that are
accessible at present colliders (be it in Higgs physics, flavor
physics or searches for physics beyond the Standard Model) are
typically rather complicated; as such, from the theoretical side, one
needs to deal with high multiplicities, and be able to go (at times
far) beyond leading-order approximations in order to adequately match
the high precision of experimental data.

As a consequence, within weak-coupling expansions, one has to consider
processes involving hundreds or even thousands of Feynman diagrams,
whose numerical evaluation sometimes requires using powerful computer
clusters with thousands of CPUs over timescales of weeks. Given this
situation, the development of both, computer tools for improving the
automation of the computations as well as advanced algorithmic
calculational methods, becomes highly relevant.

During this edition of ACAT, we have seen 18 parallel and 7 plenary
talks devoted to computations in theoretical HEP. While most of the
presented topics are closely interrelated (and in one way or the other
related to computer tools), for the sake of this summary we decided to
divide them into three main categories: computer algebra tools,
algorithms for multi-loop computations and applications to physical
processes. Given the nature of this summary, we will not attempt to do
justice to all new developments and tools in the field, but only
concentrate on the subset that has actually been discussed at the
meeting.

\section{Computer tools}

\begin{figure}
\vspace{-9mm}
\begin{center}
\includegraphics[width=0.48\textwidth]{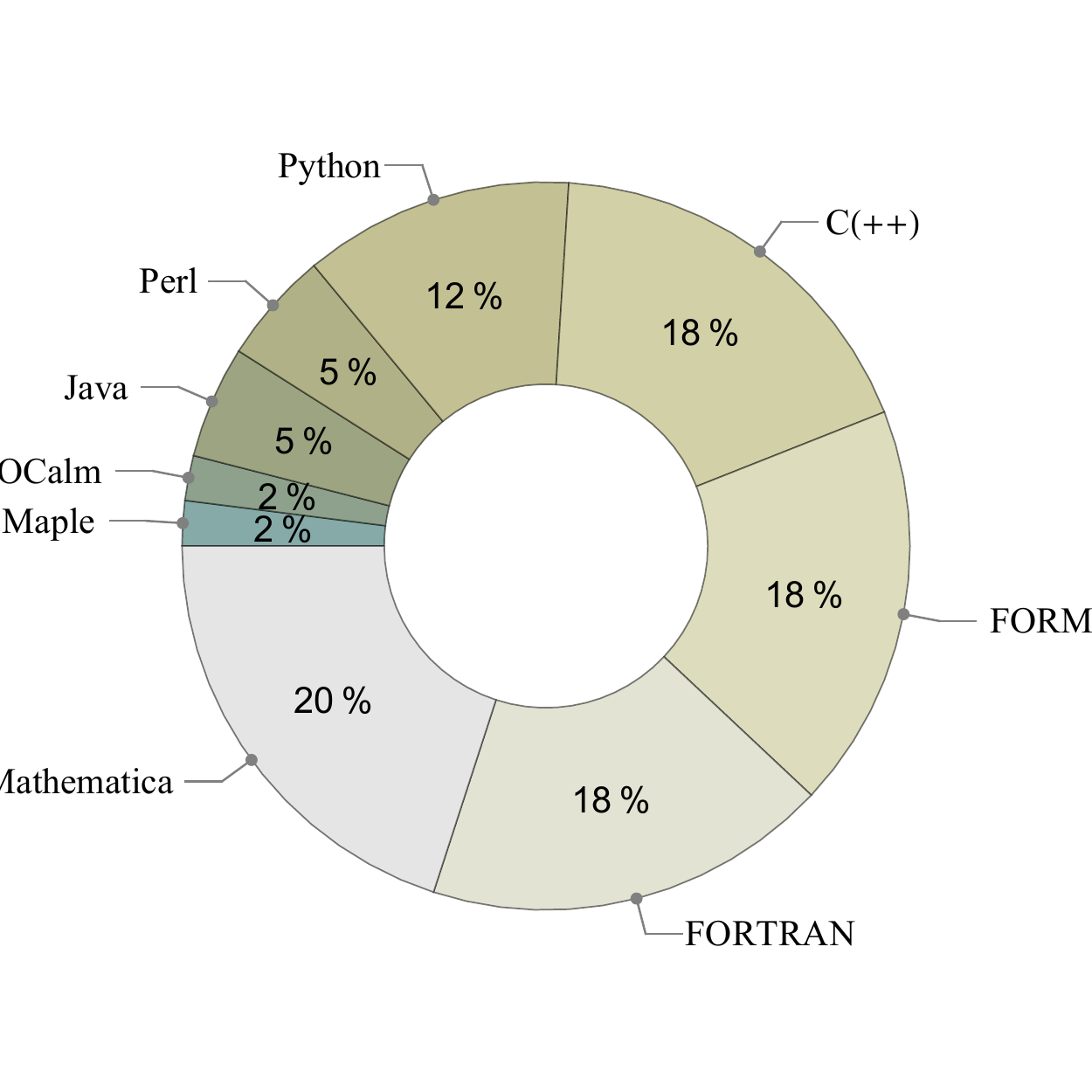}
\hspace{0.02\textwidth}
\includegraphics[width=0.48\textwidth]{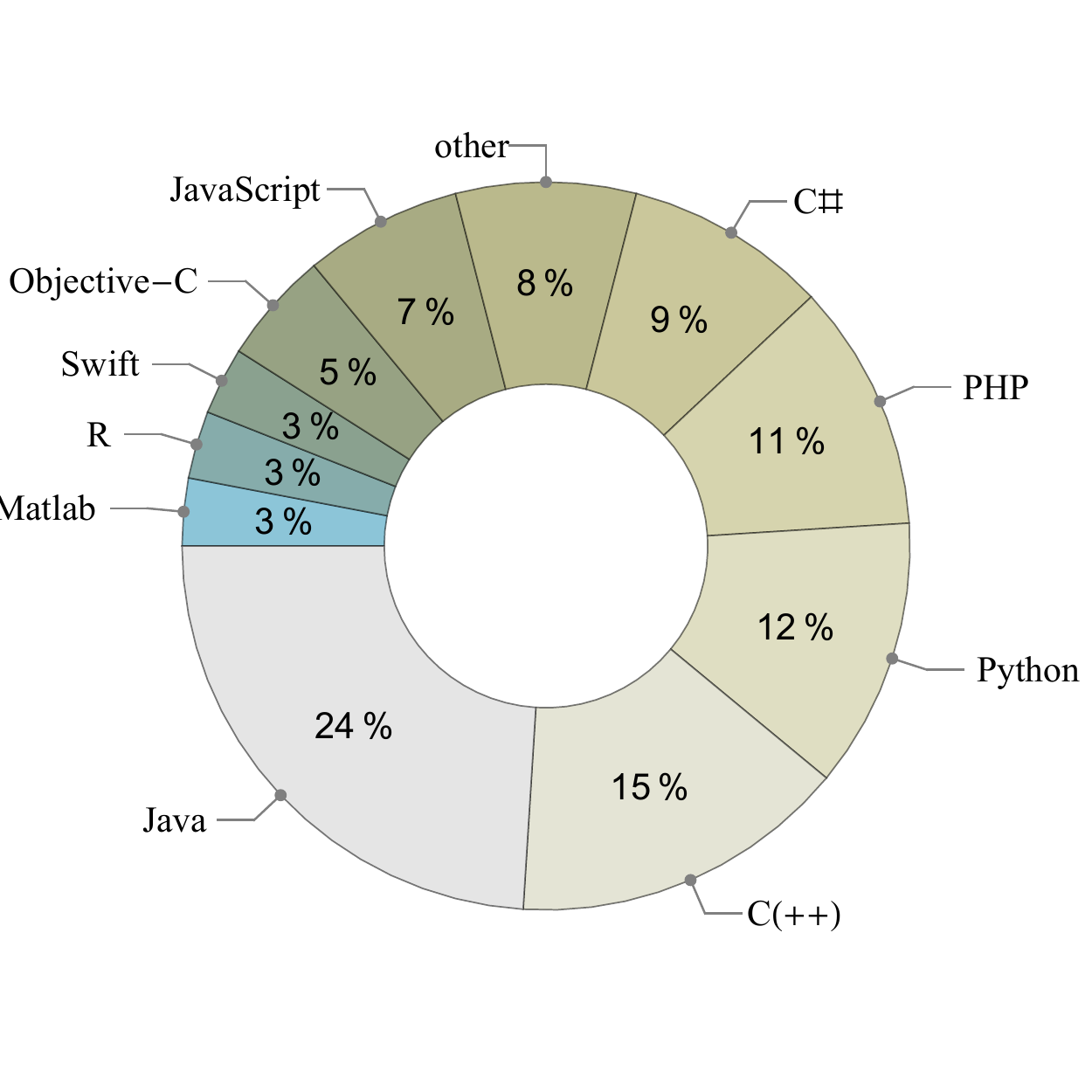}
\end{center}
\vspace{-8mm}
\caption{\label{fig_langs}The distribution of programming languages
  and tools used in the theoretical high-energy physics community
  (left panel) and in the software industry (right panel). For the
  former, we have analyzed the packages presented and used by the
  participants of the session. For the latter, data from the PYPL
  index was used \cite{PYPL}.}
\end{figure}

During the workshop, ten new computer packages or new versions of
existing packages have been presented. It is quite interesting to
analyze which programming languages and frameworks are used to code in
theoretical high energy physics research. Looking through the
different packages presented during the conference, one can find the
distribution shown in Fig.~\ref{fig_langs} (left) (for comparison, the
popularity of languages in the industry is shown in the right panel of
Fig.~\ref{fig_langs}).

We see that the \texttt{Mathematica} language \cite{Mathematica} is
the most popular. This is not surprising since many computations in
theoretical HEP are based on algebraic manipulations, and
\texttt{Mathematica} provides (besides the language itself) a wide set
of computer algebra tools coming with a very convenient user
interface.

For large-scale HEP problems, however, the performance of
\texttt{Mathematica} leaves much to be desired since it was designed
as a general-purpose computer algebra system (CAS) without any focus
on specific problems arising in HEP. Another drawback of
\texttt{Mathematica} is that it is proprietary software. These gaps
are filled by \texttt{FORM}~\cite{Vermaseren:2000nd} -- a free high
performance CAS specifically focused on the needs of HEP. Actually,
there is a consensus that today \texttt{FORM} has the highest
performance regarding computer algebra applications in HEP. It is
worth noting that, strictly speaking, \texttt{FORM} does not provide a
programming language in the full sense, but it provides enough
functionality for extension and implementation of specific routines
required by the particular physical problems.

For low-level programming and, in particular, numerical computations,
the standard languages in HEP are \texttt{C(++)} and \texttt{FORTRAN}.
While \texttt{C(++)} shares the same place in the industry and in the
physics computing community, the quite wide spread usage of
\texttt{FORTRAN} is a peculiarity of the latter.

Finally, we note that \texttt{Python} fills the place of a compact
high-level language for general-purpose programming, and as can be
seen from Fig.~\ref{fig_langs} it plays a comparable role in the
industry and physics research.

\Comment{SHTABOVENKO: FeynCalc}
Among the computer tools that are used nowadays in HEP,
\texttt{FeynCalc} \cite{Mertig:1990an} is perhaps one of the most
popular and convenient-to-use packages written on top of
\texttt{Mathematica}. The set of added features includes Dirac and
SU($N$) algebra, Feynman diagram calculation, tools for loop
computations etc. The recent release of \texttt{FeynCalc9}
\cite{Shtabovenko:2016sxi} includes many important improvements: code
quality improvements; further interoperability with external programs
(including \texttt{FeynArts} \cite{Hahn:2000kx} for diagrams
generation, \texttt{FIRE} \cite{Smirnov:2008iw} for
integration-by-parts reduction, \texttt{PackageX} \cite{Patel:2015tea}
for evaluating oneloop integrals etc.); and important new features
including tools for partial fractioning (based on the \texttt{\$Apart}
package \cite{Feng:2012iq}), or multi-loop tensor decomposition. It is
also worth mentioning that the new release shows considerable
performance improvement.

\Comment{HAHN: FormCalc}
Automation of the computational process plays an important role for
large scale problems. While it is well known how to compute each
individual ingredient of a next-to-leading order (NLO) calculation
(and one can find plenty of computer packages for each single
ingredient), it is still very convenient to have a universal tool,
which aggregates all these ingredients into a single program. Such an
approach makes the calculation more easily testable and less
error-prone. \texttt{FormCalc} \cite{Hahn:1998yk} is a well-known and
powerful tool for automatic computation of one-loop processes. It uses
\texttt{FeynArts} \cite{Hahn:2000kx} to generate diagrams for the
process of interest, then passes them to \texttt{FORM}, which performs
the $d$-dimensional Dirac algebra and other things required for
computing one-loop amplitudes, and finally it produces optimized
\texttt{FORTRAN} code, which can be compiled and run in order to
obtain the cross section and cross section distributions for the
process. A new suite of shell scripts and \texttt{Mathematica}
packages demonstrates how to flexibly use \texttt{FeynArts} and
\texttt{FormCalc} beyond one-loop, and together with other
packages. Originally constructed for the computation of the two-loop
$\mathcal{O}(\alpha_t^2)$ Higgs-mass corrections in \texttt{FeynHiggs}
\cite{Hahn:2015gaa}, the code is relatively general and may serve as a
template for similar calculations.

\Comment{ARBUZOV: SANC}
An important project towards the automation of higher-order
processes is the computer system \texttt{SANC}
\cite{Andonov:2004hi}, for which recent developments were presented in
the plenary talk of Andrej Arbuzov. \texttt{SANC} has its origins in
the \texttt{ZFITTER} program \cite{Arbuzov:2005ma} for the calculation
of fermion pair production and radiative corrections at high energy
$e^+e^-$ colliders, but it additionally supports QCD processes
(including convolution with partonic distributions) and both QCD and
EW corrections at the one-loop precision level. The list of processes
implemented in \texttt{SANC} includes Drell-Yan processes (there
was recently an important update \cite{Arbuzov:2015yja} including a
systematic treatment of the photon-induced contribution in
proton-proton collisions and electroweak corrections beyond NLO
approximation), associated Higgs and gauge boson production,
and single-top quark production in $s$- and $t$-channels
\cite{Bondarenko:2013nu}.

\Comment{POSLAVSKY: Redberry}
The new computer algebra system \texttt{Redberry}
\cite{Bolotin:2013qgr} is focused specifically on the applications in
high energy physics and gravity. The main feature of this system is
that it considers all mathematical objects that arise in HEP
(ordinary scalars, vectors, spinors, matrices and in general arbitrary
tensors) in a uniform and robust way. The graph-theoretical approach
implemented in \texttt{Redberry} allows to achieve unmatched
performance of operations involving symbolical
tensors. While \texttt{Redberry} comes with an extensive set of programming features which make it easily extendable, out of the box it provides a range of common tools for performing computations in HEP such as Dirac \& SU(N) algebra, simplifications with spinors, or calculation of one-loop counterterms in curved space-time, amongst others. Recent developments include derivation of Feynman rules from a given Lagrangian, and the computation of NLO processes.

\Comment{REUTER: VMs}
As we have mentioned above, in order to match experimental precision
it sometimes needed to consider high multiplicity processes, whose
evaluation requires to run on large computer clusters for days or even
weeks. For some important processes, the standard numerical approach
quickly becomes infeasible. For example, the size of a file containing
the matrix element for $2 \to 6$ gluon scattering written in a
\texttt{FORTRAN} code is about 2 gigabytes. It is difficult to compile
file which are that large in reasonable time. In such cases, some kind
of virtual machine (VM) can prove very useful. Instead of writing the
analytical expression for the amplitude in some language like
\texttt{C(++)} or \texttt{FORTRAN} and compile it, one can translate
it into a binary format, which then can be interpreted by the VM and
calculated just-in-time
(so-called JIT compilation) -- this means that
the actual translation to machine instructions done during execution of a program (at run time) rather than prior to execution. This paradigm is
implemented as an extension \cite{Nejad:2014sqa} for the Optimized
Matrix Element Generator O'Mega \cite{Moretti:2001zz}. The implemented
technique is very simple and can be naturally parallelized on several
computational cores or even on GPUs. The performance shown using this
approach is comparable to compiled code, and scales favorably for
extremely high multiplicity processes.

\Comment{REUTER: \texttt{Whizard}}
The computation of NLO corrections for multi-leg processes requires a
large degree of automation in the generation and computation of the
amplitudes and the phase space. \texttt{Whizard} is one of the several
tools able to perform this task. Despite being used mainly for lepton
collider predictions, it can compute predictions also for
proton-proton collisions. The new release (v.2.2.8) features an
implementation of the Binoth Les Houches Accord (BLHA), to be able to
interface with codes computing one-loop corrections, an automatic
generation of subtraction terms based on the formalism developed by
Kunszt, Frixione and Signer~\cite{Frixione:1995ms} and an
implementation of the \texttt{POWHEG} matching~\cite{Frixione:2007vw}
of fixed order NLO computations with parton shower
algorithms. Furthermore a new phase space remapping was introduced for
a correct treatment of resonances in NLO computations. The release of
the new version was accompanied by validation examples based on
several processes in electron-positron annihilation~\cite{acatReuter}.

\Comment{HEINRICH: \texttt{SecDec 3.0}}
Another tool used for the computation of NLO corrections for several
LHC processes is \texttt{GoSam}~\cite{Cullen:2011ac,Cullen:2014yla}. A
concrete example is mentioned later in
Section~\ref{sec:pheno}. \texttt{GoSam} provides a framework for the
automatic computation of one-loop amplitudes and is now being upgraded
to be able to deal also with two and higher loop amplitudes. The basic
idea is to use the interface to \texttt{QGRAF}~\cite{Nogueira:1991ex}
and \texttt{FORM}, already present in \texttt{GoSam}, to generate the
amplitudes and reduce them to a linear combination of master
integrals, which can than be computed with another external tool. One
of these tools in the code \texttt{SecDec}, of which version 3.0 was
recently released~\cite{Borowka:2015mxa}. \texttt{SecDec} can compute
multi-loop master integrals using the method of sector decomposition,
by producing a Laurent expansion of the integral and computing the
coefficients of the expansion numerically. The new version has a
better user interface and can be run also on clusters. Furthermore,
besides the several improvements in the possible reduction strategies,
it allows to compute integrals containing propagators with zero or
negative power and also complex masses. All these new features are
currently extensively tested on practical examples and in particular
for the computation of NLO QCD corrections to double Higgs boson
production in gluon-gluon fusion~\cite{acatHeinrich}.

\section{Techniques for (multi-)loop calculations}
\label{sec:multiloop}

One of the main motivations behind the most recent developments on
advanced techniques for Feynman-diagrammatic computations is the need
of precise QCD predictions. These are going to be crucial to make the
best out of the large amount of data, which will be collected during
the LHC Run2. As example we remind that, in some phase-space regions,
NNLO QCD effects can be of the order of 20\% (such as e.g. in diboson
production). In this respect, increasing the precision of theory
predictions by including more loops and/or more legs almost always
leads to an improved accuracy. As reflected in many of the talks, we
have witnessed a number of recent developments which pushed the
boundaries of what can be done on the multi-loop front further, both
analytically as well as numerically.

These developments are driven by many new ideas, of which we had
interesting overviews in the plenaries. Some of the key advances
include: the integration-by-parts (IBP) over finite fields, a broad
attack on the bottlenecks of Feynman integral
reduction~\cite{vonManteuffel:2014ixa}, the employment of on-shell
methods to automate high-multiplicity QCD amplitude
computations~\cite{Badger:2015lda}; the usage of hyperlogarithms in an
attempt to classify the functional content of a wide class of
integrals~\cite{Panzer:2015ida} and an elliptic generalization of the
ever-present multiple polylogarithms (MPLs), to allow for insight into
new function classes lurking behind Feynman
diagrams~\cite{Adams:2014vja}.

As a general and uniting theme, most of the research programs devoted
to multi-loop techniques are pushing towards highly algorithmic
methods, due to the high level of automation needed in modern
perturbative quantum field theory computation. To this end, many
(public as well as private) codes, ideas and results have been
discussed in the sessions. We will give a brief account of them below.

\Comment{BADGER: QCD amplitudes -- plenary}
For long time one-loop calculations have been the bottleneck in the
computation of NLO corrections. In the last decade, the development of
systematic reduction algorithms and automated techniques for the
computation of amplitudes allowed to make a big step forward. This led
to tools which can automatically perform NLO QCD and Electroweak
computations. The same ideas are now being applied and improved also
beyond one-loop and many efforts are put in automation for
NNLO. Besides the recent improvement in the computation of subtraction
terms, the main bottleneck at NNLO is again given by the two
loop amplitudes. Despite the several recent results on $2\to2$
processes, only few things are known about two loop amplitudes for
$2\to3$ processes, which became recently a very active field of
research~\cite{Badger:2015lda}.  A very nice overview of the latest
development was given in the plenary talk by Simon
Badger.

\Comment{BOGNER: elliptic polylogs -- plenary}
As has become clear in recent studies of the algebraic structure of
Feynman integrals, many of them can be expressed in terms of the above-mentioned MPLs.
There are, however, a number of well-known exceptions, most notably
when studying electro-weak physics, QCD, or even massless N=4
Super-Yang-Mills theories. As Christian Bogner has reviewed in his
plenary talk, the former type of integrals have by now been completely
understood in terms of their iterated structure, a development that
has led to software (such as \texttt{HyperInt} \cite{Panzer:2014caa}
and \texttt{MPL} \cite{Bogner:2015nda}) which analytically solve a
large class of Feynman integrals that respect a certain condition (the
so-called linear reducibility of their associated graph
polynomials). Going beyond linear reducibility (the two-loop massive
sunset integral being a simple example), he furthermore presented a
remarkable and new class of functions that generalize the MPLs, the
elliptic polylogarithms \cite{acatBogner}, which harbor promising
potential to become standard building blocks, once their structure is
understood in more detail.

\Comment{DAVYDYCHEV: geometry}
In order to predict the functional content of the special case of
one-loop, $N$-leg Feynman integrals, it proves useful to switch to a
geometric interpretation of their Feynman parameters, mapping
integrals onto volumes. The minimal set of independent kinematic
invariants and masses that the function depends on corresponds to
objects such as a 2-dimensional triangle (for the 2-point function),
the 3d tetrahedron (at 3pt), a 4d simplex (at 4pt)
etc.~\cite{acatDavydychev}. From this (non-Euclidean) geometric
intuition, it is possible to derive additional relations among the
parameters, essentially reducing the number of variables from $N^2$ to
$N$, which in facts severely constrains the functional forms. An
all-$N$ proof of this reduction of complexity, as well as a relation
to other geometric constructions, such as the amplituhedron, is still
open.

\Comment{KONDRASHUK: new basis}
It is sometimes possible to use iterative structures of a specific set
of Feynman integrals in order to obtain results for high (or even all)
loop orders. As an example, 3-point ladder-type functions provide such
a set, relevant in planar sectors of gauge theory. In particular,
utilizing repeated transformations between momentum- and dual space
when the propagator powers satisfy specific relations
\cite{Usyukina:1993ch}, one can systematically reduce the number of
rungs, eventually mapping on sums of lower-loop integrals
\cite{acatKondrashuk}.  It would be interesting to check this method
against the above-mentioned criterion of linear reducibility.

\Comment{MANTEUFFEL:  multiloop methods -- plenary}
Looking at Feynman integrals from a more general perspective, one can
view them as forming a linear vector space, since the well-known IBP
relations form linear relationships between them. The problem
of performing perturbation theory in an efficient way can then be
regarded as choosing the most convenient basis. To this end, Andreas
von Manteuffel highlighted three methods of different maturity level
in his plenary talk. First, the master integrals required for
the full NNLO QCD corrections to diboson production at LHC have been 
solved via a linear system of first order differential
equations in invariants, mapping onto MPLs after a suitable basis
change, which allow to construct optimized functions for fast and
stable numerics by requiring absence of spurious denominators. 
Second, it was discussed how the problem of divergences in Feynman 
parameter integrals (which obstruct direct $\varepsilon$\/-expansions 
on the integrand level) can be overcome by changing to a basis of quasi-finite 
integrals and perform a direct expansion and integration
circumventing sector decomposition or analytic regularization 
\cite{vonManteuffel:2014qoa}. Third, observing that the polynomial GCD
operations needed for the rational functions in large IBP reduction
problems consume most of the CPU time, significant
speedups may be obtained by switching to modular arithmetic and substituting 
variables with integers, reducing modulo finite prime fields, and using 
rational reconstruction (Hensel lifting) to reconstruct the original answer
\cite{vonManteuffel:2014ixa}.

\Comment{HOFF: TopoID}
Turning towards the classification of Feynman integral families and
topologies, a new \texttt{Mathematica} package \texttt{TopoID} has
been presented \cite{acatHoff}. It can be used in various ways, such
as in preprocessing IBP identities or optimizing integral families,
and works on the level of graph polynomials, providing a unique
ordering that allows for unique and efficient topology mapping. As an
application and first check, \texttt{TopoID} has been successfully
used in evaluating N$^3$LO corrections to the Higgs boson production
cross section \cite{Anzai:2015wma}, while ultimately aiming for
e.g. optimizing the evaluation of 5-loop propagators, with
applications to SM parameters such as the 5-loop QCD Beta function.

\Comment{SCHRODER: 5loop tadpoles}
Pushing the loop frontier, a study of the class of five-loop fully
massive vacuum diagrams has been presented \cite{acatSchroder}. This
specific class finds applications in e.g.\ QCD thermodynamics and
anomalous dimensions.  Using IBP methods to identify the complete set
of master integrals, those have been evaluated numerically to high
precision, using IBP, difference equations, and factorial series. A
parallelized (private) \texttt{C++} implementation employs
\texttt{Fermat} for polynomial algebra, and implements substantial
fine-tuning of the classic Laporta approach \cite{Laporta:2001dd}. Due
to the diverse nature of envisaged applications, the corresponding
code \texttt{TIDE} is able to optimize large
systems of difference and recursion relations for
$\varepsilon$\/-expansions around 3 and 4 dimensions.

\Comment{KOMPANIETS: 6loop}
The loop-level record at this meeting, however, has been set by
Mikhail Kompaniets who reported on analytic results for the six-loop
beta function in scalar theory \cite{Batkovich:2016jus}. Massively
profiting from treating $\varphi^4$ theory where each Feynman diagram
corresponds to a single integral only (as opposed to orders of
magnitudes more in gauge theories), the presented strategy restricts
to the 3-vertex-free subset of 6-loop massless propagators without
dots. Relying on IBP paired with a (private) R$^*$ implementation for
systematically disentangling UV- and IR-divergences, the 6-loop
integrations can ultimately be performed in terms of hyperlogarithms
(see above), to arrive at an analytic textbook-quality result in terms
of multiple Zeta values.

\Comment{UEDA: Forcer}
For specific classes of Feynman integrals, it is possible to build a
package that performs an exact reduction and solution of any given
member, hence providing a method for exact evaluation of that
class. For massless two-point functions, this has been achieved at the
three-loop (\texttt{Mincer} \cite{Gorishnii:1989gt}) and most recently
on the four-loop level \cite{acatUeda}. Due to the large tree of
different integral sectors, the latter development has been made
possible through a high level of automation (``write a small program
that produces a large (41k lines) one''), incorporates new IBP tricks
such as the exact diamond rule \cite{Ruijl:2015aca}, and culminated in
the \texttt{FORM} code \texttt{Forcer} that has already been
successfully tested on a number of problems, and which will enable the
authors to evaluate e.g.\ the $N=6$ Mellin moments of 4-loop splitting
functions. Looking ahead, the automated derivation of exact reduction
rules might enable even a `5-loop Mincer' package.

\Comment{KATO: Direct computation method}
Another way to compute loop integrals numerically, is to explicitly
make use of the Feynman $\epsilon$-prescription computing the integral
for a finite value of $\epsilon$ and than extrapolate to the limit for
$\epsilon\to 0$. This so called \textit{direct computation method},
presented by K.~Kato, needs high performance computing facilities, but
allows to compute diagrams at several loops in a systematic
way. Examples were shown for 2-, 3- and 4-loop diagrams.

\section{Phenomenological applications}
\label{sec:pheno}

The complexity and variety of theoretical HEP computations is such
that not all computations can directly be used for phenomenology. This
was the case for several computations described in the previous
sections, which are fundamental building blocks for future
phenomenological applications, but whose outcome can currently not yet
be compared with experimental data. In this section however we present
some results which have direct impact on phenomenology.

\Comment{KATAEV: running masses}
The important problem of the relation between pole and running masses
of heavy quarks is still a very active research field. To date, we
have some known results for these relations up to four loops, part of
which are analytical, while the complete result is known numerically
only and has been obtained very recently \cite{Marquard:2015qpa}. This
numerical result is known for a fixed number of flavors only. The
authors of Ref.~\cite{Kataev:2015gvt} used a least squares method in
order to extract the full dependence of the relation between heavy
quark pole- and running mass on the number of flavors. They show that
results for pole masses of heavy quarks form an asymptotic series,
which is a well-known feature of perturbative series expansions. For
the bottom quark, the asymptotic nature arise only at the four-loop
level, while for the top quark even up to four loops inclusive, the
perturbation series converges. So, for the $t$ quark it is natural to
take into account these corrections. As of today, experimental
uncertainties in the top mass are larger than the corresponding theory
errors, but this situation might be reversed in the nearest future, in
which case the presented results come to play an important
role. Additionally, the least squares method used by the authors
allows to estimate theoretical uncertainties which is quite
informative when using numerical methods. These uncertainties turn out
to be three times larger than the numerical errors obtained in
\cite{Marquard:2015qpa}; however, the authors note that their
uncertainties may be overestimated, an issue that can be resolved by
the direct calculation of the full analytical result.

\Comment{CVETIC: Bottom mass from quarkonium}
Another important aspect related to quark masses, is their
experimental determination. The $\mathrm{\overline{MS}}$ bottom-quark mass can
for example be determined precisely from $\mathrm{Y}(1S)$, which is
the lowest energy $b\bar{b}$-quarkonium state. The mass of the
quarkonium state can be written as a sum of the bottom quark pole
mass and the binding energy. These quantities however suffer from
renormalon ambiguity. This issue can be solved by expanding them in powers
of the ratios over the $\mathrm{\overline{MS}}$ bottom-quark mass and by
subtracting the leading renormalon extracted from the static
potential. With this technique it was recently possible to determine
the value of the $\mathrm{\overline{MS}}$ bottom-quark mass to percent
accuracy~\cite{Ayala:2014yxa}.

\Comment{BOUGHEZAL}
The recent discovery of the long-sought-after Higgs boson by the ATLAS
and CMS collaborations has ushered in an era of precision measurements
to determine the nature of the new particle. The recent computation of
Higgs boson production in gluon-gluon fusion at N$^{3}$LO
accuracy~\cite{Anastasiou:2015ema}, or the computation of the
differential NNLO QCD corrections of $\mathrm{H}\!+\!1$
jet~\cite{Boughezal:2015dra}, $\mathrm{Z}\!+\!1$
jet~\cite{Boughezal:2015ded} or $\mathrm{W}\!+\!1$
jet~\cite{Boughezal:2015dva} are just few examples. The fact that it
was recently possible to compute so many processes at NNLO accuracy in
relatively short time relies on better automation, on improved
algorithms (an example are the new subtraction methods developed
recently~\cite{Gaunt:2015pea}), and on the smart combination of
parallelization algorithms which allow to perform this calculations on
large high performance computing facilities, obtaining the results
within few hours of running.

\Comment{LUISONI: H+jets in gluon-gluon fusion}
Another important aspect to keep in mind when performing such
CPU-expensive computations is the question of storing results in the
most flexible way possible. A possibility which became popular
recently for NLO computations, is to store events in form of
\texttt{Root} N-Tuples~\cite{Bern:2013zja}. This allows to perform the most
computationally expensive part of the calculation only once and to
re-run the phenomenological analysis varying cuts, parton distribution
functions or scales several time in a fast way. An example of a possible
study that can be performed with N-Tuples was presented during the
conference for the NLO QCD corrections to
$\mathrm{H}\!+\!1,\;2,\;\mathrm{and}\;3$ jets~\cite{acatLuisoni}.

\Comment{KULAGIN: nPDFs}
Hadronic parton distribution functions (PDFs) determine the leading
contributions to the cross sections of various hard processes. This
topic was covered in the plenary talk given by Sergei Kulagin. The
PDFs of hadrons determine the leading contributions to the cross
sections of various hard processes. The deep-inelastic scattering
(DIS) and Drell-Yan (DY) experiments with nuclei demonstrated
significant nuclear effects on the parton level. These observations
rule out the naive picture of the nucleus as a system of quasi-free
nucleons and indicate that the nuclear environment plays an important
role even at energies and momenta much higher than those involved in
typical nuclear ground state processes. Aiming at a better
understanding of nuclear parton distributions, a detailed
semi-microscopic model of nuclear PDFs has been developed. It includes
the QCD treatment of nucleon structure functions, and addresses a
number of nuclear effects such as shadowing, Fermi motion and nuclear
binding, nuclear pion and off-shell corrections to bound nucleon
structure functions
\cite{Kulagin:2004ie,Kulagin:2010gd,Kulagin:2014vsa}. Using this approach,
the authors were able to reproduce data on nuclear effects in DIS and 
DY experiments. Predictions for the neutrino-nuclear cross sections 
have also been discussed.

\Comment{ALEKSEJEVS: Moller scattering}
Automation and precision is needed also in searches for physics Beyond
the Standard Model (BSM). This is for example the case when performing
very precise computations of Standard Model process to search for
indirect deviations from the predictions. An example presented during
the conference refers to the first computation of NNLO corrections,
namely the interference of one-loop contributions, to the polarized
M{\o}ller scattering~\cite{acatAleksejevs}. It is also possible to
consider the effect of considering a further $Z^{\prime}$ in the
theoretical predictions and produce precise exclusion limits.

\Comment{WALTENBERGER: smodels}
Exclusion limits are however also produced by several experimental
analysis. The question can therefore also be reversed: how can the BSM
exclusion limits determined by the LHC experiment be systematically
exploited to discriminate between the many possible BSM extensions? A
possible answer is given by the \texttt{SModels}
project~\cite{Kraml:2013mwa}. The key idea is to devise a formal
language to describe the LHC results collected into a database, which
can then be matched to the decomposed spectrum of each given
model. This allows to quickly check if the experimental results
already exclude a given model or whether there is still room for it to
be realized in nature.

\begin{figure}
\begin{center}
\includegraphics[width=0.8\textwidth]{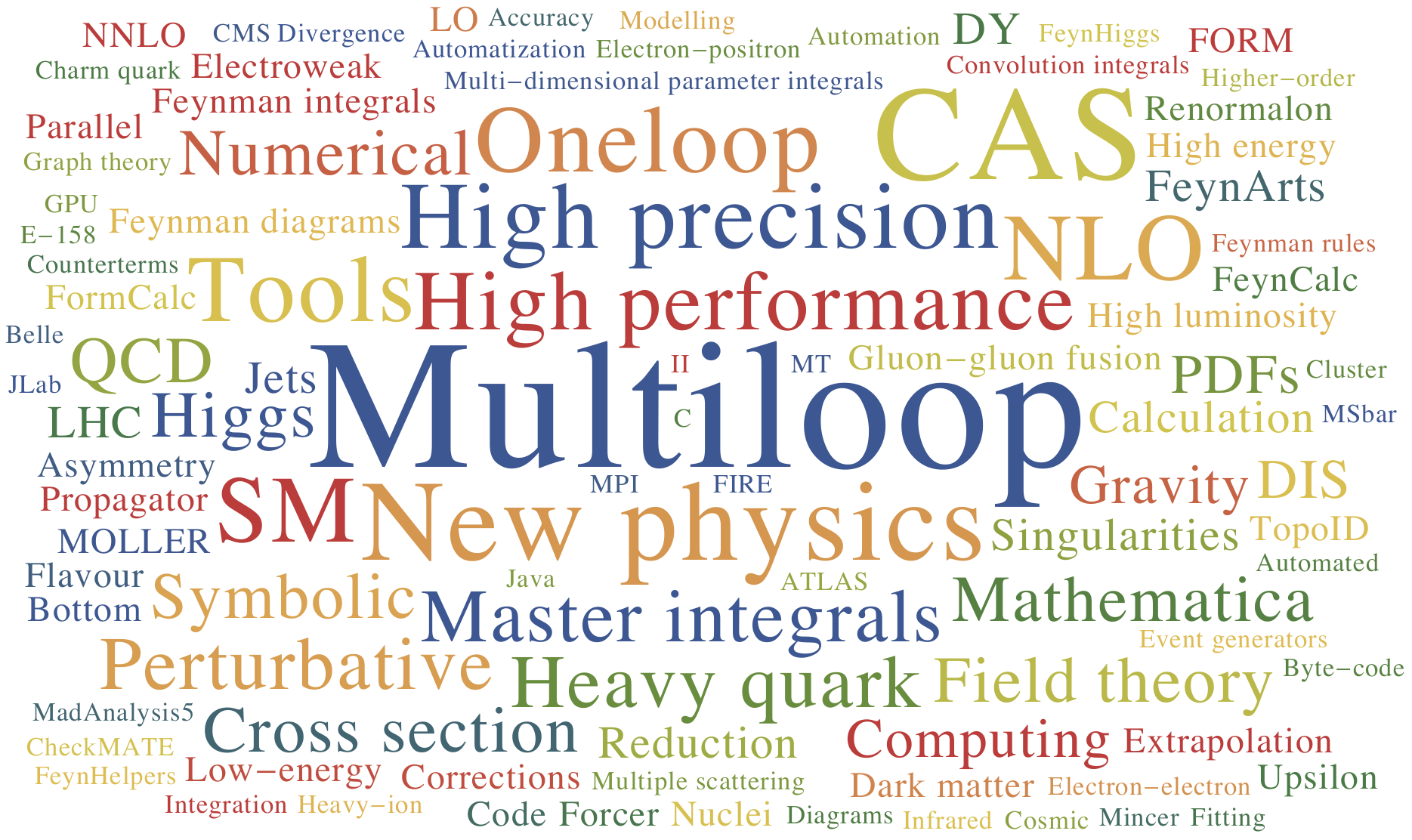}
\end{center}
\caption{\label{fig:wordCloud}Typical keywords of Track 3 presentations.}
\end{figure}

\Comment{CURE:Astrophysics}
Rotational speed is an important physical parameter of stars: knowing
the distribution of stellar rotational velocities is essential for
understanding stellar evolution. However, rotational speed cannot be
measured directly, but is given by the convolution between rotational
speed and the sine of the inclination angle -- $v \sin(\alpha)$. The
problem of deconvolving stellar rotational speeds was discussed in the
plenary by Michel Cure. The problem itself can be described via a
Fredholm integral of the first kind. The new method \cite{Cure} to
deconvolve this inverse problem and to obtain the cumulative
distribution function for stellar rotational velocities is based on
the work of Chandrasekhar and M\"unch \cite{Chandrasekhar}. The
proposed method can be also applied to the distribution of extrasolar
planets and brown dwarfs (in binary systems) \cite{Cure15}. For stars in a cluster, where all members
are gravitationally bounded, the standard assumption that rotational
axes are uniformly distributed over the sphere is questionable. On the
basis of the proposed techniques, a simple approach to model this
anisotropy of rotational axes was developed, opening up the
possibility to `disentangle' simultaneously both the rotational speed
distribution and the orientation of rotational axes.

\section{Conclusions}

We leave the reader with the word cloud presented in
Fig.~\ref{fig:wordCloud}, which attempts to represent the main
keywords of all Track 3 talks, as drawn from the actual presentations.

\ack

We would like to thank the organizers of the ACAT 2016 workshop,
especially Luis Salinas and Claudio Torres, who took the major part of
the organization. Our thanks extend to all involved staff of the
Technical University Federico Santa Maria who provided a very
comfortable atmosphere during the workshop. We would like to thank
Andrei Kataev, Tord Riemann and Igor Kondrashuk for valuable
discussions. The work of S.P.\ was supported by the Chilean Center for
Mathematical Modeling (CMM) and partially by the RFBR grant
\#16-32-60017 and by the FAIR-Russia Research
Center. Y.S.\ acknowledges support from DFG grant SCHR 993/1, FONDECYT
project 1151281 and UBB project GI-152609/VC.

\section*{References}
\bibliography{summaryV5}

\end{document}